\documentclass[CEJP,DVI]{cej} % use DVI command to enable LaTeX driver
\usepackage{layout}
\usepackage{amsmath}
\usepackage{textcomp}
\usepackage{hyperref}
\articletype{Research Article}

\title{Equation of state and initial temperature of quark gluon plasma at RHIC}

\author{M\'at\'e~Csan\'ad\inst{1}\email{csanad@elte.hu},
             Imre~M\'ajer\inst{1}}
\institute{\inst{1}E\"otv\"os University,
 Department of Atomic Physics, P\'azm\'any P\'eter s. 1/A, H-1117 Budapest, Hungary}

\abstract{
In gold-gold collisions of the Relativistic Heavy Ion Collider a perfect fluid,
the strongly interacting quark gluon plasma (sQGP) is created.
The time evolution of this fluid can be described by hydrodynamical models. After an expansion, hadrons are
created during the freeze-out. Their distribution reveals information about the final state.
To investigate the time evolution one needs to analyze penetrating probes: e.g. direct
photon observables.
In this paper we analyze a 1+3 dimensional solution of relativistic hydrodynamics.
We calculate momentum distribution, azimuthal asymmetry and momentum correlations of direct photons. Based on earlier
fits to hadronic spectra, we compare photon calculations to measurements
to determine the equation of state and the initial temperature of sQGP. We find that the initial
temperature in the center of the fireball is 507$\pm$12 MeV, while for the sound speed we
get $c_s$=0.36$\pm$0.02. We also estimate a systematic error of these results. We find that the measured
azimuthal asymmetry is also not incompatible with this model, and predict a photon
source that is significantly larger in the \emph{out} direction than in the \emph{side} direction.
}

\begin{document}

\maketitle
\section{Introduction}
In the last several years it has been revealed that the matter produced in the collisions of the Relativistic Heavy Ion Collider (RHIC)
is a nearly perfect fluid~\cite{Lacey:2006bc}, i.e.\ it can be described with perfect fluid hydrodynamics. There was a long search for
exact hydrodynamic models (solutions of the partial differential equations of hydrodynamics) and several models proved to be
applicable~\cite{Landau:1953gs,Khalatnikov:1954aa,Belenkij:1956cd,Hwa:1974gn,Chiu:1975hw,Bjorken:1982qr,Biro:2000nj,Csorgo:2006ax,Bialas:2007iu}. Many solve the equations of hydrodynamics numerically, which has the advantage of having arbitrary initial conditions.
However, it is possible to find analytic solutions as well, see the above cited references for example. These solutions will have a couple of
parameters, the optimal value of which can then be determined by fitting theoretical results to data. Among these models there are
non-relativistic 1+3 dimensional ones,  as well as 1+1 dimensional relativistic models - but not many 1+3 dimensional \emph{and} relativistic
models were tested yet. In this paper we extract photon observables from the relativistic, ellipsoidally symmetric model of
 Ref.~\cite{Csorgo:2003ry}. Hadronic observables were calculated in Ref.~\cite{Csanad:2009wc}. Here we calculate transverse momentum distribution and azimuthal asymmetry (elliptic flow) of direct photons. We also calculate Bose-Einstein correlations and their widths, called
HBT radii.

For the direct photon calculations, the framework used in this paper has to be made very clear. Our key assumption is, that
even though direct photons may not be thermalized in the strongly interacting plasma (as their mean free path may be on the order
of the size of the fireball), but the radiation itself is thermal. Thus the phase-space distribution of the photons is characterized by the
temperature of the medium (at a given space-time cell), while the expansion of the fireball also effects the observed spectrum. This is a
 macroscopic model, and in the following we will calculate photon observables from it and compare to RHIC data. The most important
assumption is, that the spectrum of direct photons is thermal because macroscopically, the photon radiation is thermal. Our tool will be an exact solution of relativistic hydrodynamics, applied to the expanding medium, the strongly interacting plasma created in Au+Au collisions
at RHIC.

\section{Perfect fluid hydrodynamics}
Perfect fluid hydrodynamics is based on local conservation of a conserved charge ($n$) and energy-momentum ($T^{\mu\nu}$):
\begin{align}
\label{e:rhd}
		\partial{_\mu}(nu^\mu)&=0,\\
		\partial{_\mu}T^{\mu\nu}&=0,
\end{align}
where $u^\mu$ is the flow field in the fluid. These quantities here refer to the strongly interacting plasma
created at RHIC, i.e.\ the equations dealing with its temperature, flow etc. The fluid is perfect if the energy-momentum
tensor is diagonal in the local rest frame, i.e.\  viscosity and heat conduction are negligible.
Thus $T^{\mu\nu}$ is chosen as
\begin{align}
T^{\mu\nu}=(\epsilon+p)u^\mu u^\nu-pg^{\mu\nu},
\end{align}
where $\epsilon$ is energy density, $p$ is pressure and $g^{\mu\nu}$ is the metric tensor, diag(-1,1,1,1). Note also that
in this paper, $x^{\mu} = (t, r_x, r_y, r_z)$ is a given point in space-time, $\partial_{\mu} = \partial/\partial x^{\mu}$
is the derivative versus space time, while $p^{\mu} = (E, p_x, p_y, p_z)$ is the four-momentum.

The conservation equations are closed by the equation of state (EoS), which gives the
 relationship between energy density $\epsilon$ and pressure $p$. Typically $\epsilon = \kappa p$ is chosen,
where the proportionality ``constant'' $\kappa$ may depend on temperature $T$. Note also
that $\kappa$ equals $1/c_s^2$, with $c_s$ being the speed of sound.
Temperature in turn is connected to density $n$ and pressure $p$ via $p=nT$. In some solutions
(as also on the analyzed one) a bag constant $B$ can be introduced (this is however not favored
by the data, because no first order phase transition is seen in high energy heavy ion
collisions, these are in the cross-over regime of the QCD phase-map~\cite{Fodor:2009ax}).
The exact, analytic result for hydrodynamic models is, that the hadronic observables do not
depend on the initial state or the dynamical equations separately, just through the final
state~\cite{Csanad:2009sk}. But if we fix the final state from hadronic data, we can determine
initial state parameters from direct photon spectra.

Even though many solve the above equations numerically, there are only a few exact solutions for these equations. One (and 
historically the first) is the implicit solution discovered more than 50 years
ago by Landau and Khalatnikov~\cite{Landau:1953gs,Khalatnikov:1954aa,Belenkij:1956cd}. This
is a 1+1 dimensional solution, and has realistic properties: it
describes a 1+1 dimensional expansion, does not lack acceleration
and predicts an approximately Gaussian rapidity distribution.
Another renowned solution of relativistic hydrodynamics was found by
Hwa and Bjorken~\cite{Hwa:1974gn,Chiu:1975hw,Bjorken:1982qr}:
it is simple, 1+1 dimensional, explicit and exact, but
accelerationless. It is boost-invariant in its
original form. Boost invariance is however incompatible with data
from RHIC, so the solution fails to describe the data, it still can be used
to estimate the energy density in high energy heavy ion collisions.
Important are solutions~\cite{Csorgo:2006ax,Bialas:2007iu} which are
explicit and describe a relativistic acceleration, i.e. combine the
properties of the Landau-Khalatnikow and the Hwa-Bjorken models. With these models
one can have an advanced estimate on the energy density, but investigation of transverse
dynamics is not possible.

There was only one 1+3 dimensional relativistic solution investigated: the solution in Ref.~\cite{Csorgo:2003ry}.
Hadronic observables from this solution were computed and compared to data in Ref.~\cite{Csanad:2009wc}. Present
paper calculates thermal photon observables from this realistic 1+3 dimensional model and compares them to data for
the first time. Our method is different from numerical calculations: here one can determine the best values of the
parameters of the solution by fitting the analytic model results to data.

\section{The analyzed solution}
The analyzed solution~\cite{Csorgo:2003ry} assumes self-similarity and ellipsoidal symmetry, as described in Ref.~\cite{Csanad:2009wc}.
The ellipsoidal symmetry means that at a given proper time the thermodynamical quantities are
constant on the surface of expanding ellipsoids.
The ellipsoids are given by constant values of the scale variable $s$, defined as
\begin{align}
s=\frac{r_x^2}{X(t)^2}+\frac{r_y^2}{Y(t)^2}+\frac{r_z^2}{Z(t)^2},
\end{align}
where $X(t)$, $Y(t)$, and $Z(t)$ are time dependent scale parameters (axes of the $s=1$ ellipsoid),
only depending on the time. Spatial coordinates are $r_x$, $r_y$, and $r_z$.
The velocity-field (of the fireball created in RHIC collisions) is described by a Hubble-type expansion:
\begin{align}
u^\mu (x) = \gamma \left(1, \frac{\dot X(t)}{X(t)}r_x, \frac{\dot Y(t)}{Y(t)}r_y, \frac{\dot Z(t)}{Z(t)}r_z\right),
\end{align}
where $x$ means the four-vector $(t,r_x,r_y,r_z)$, and $\dot X(t) = dX(t)/dt$, similarly for $Y$ and $Z$.
The $\dot X(t)=\dot X_0$, $\dot Y(t)=\dot X_0$, $\dot Z(t)=\dot X_0$ (i.e. all are constant) criteria must
be fulfilled, ie.\ the solution is accelerationless. This is a drawback of this model.

The temperature distribution $T(x)$ is
\begin{align}
T(x)=T_0\left(\frac{\tau_0}{\tau}\right)^{3/\kappa} \frac{1}{\nu(s)},\label{e:temp}
\end{align}
where $\tau$ is the proper time, $s$ is the above scaling variable, $\nu(s)$ is an arbitrary function, while $T_0=T|_{s=0,\tau=\tau_0}$,
and $\tau_0$ is an arbitrary proper time, but practically we choose it to be the time of the freeze-out, thus $T_0$ is the central freeze-out temperature.
The function $\nu(s)$ is chosen as $\nu(s)=\exp(-bs/2)$
where $b$ is proportional to the temperature gradient. If the fireball is the hottest in the center, then $b<0$.
If there is a conserved charge in the system e.g.\ the baryon number density, then charge number density $n(x)$ can
be utilized in the solution. As described in Refs.~\cite{Csorgo:2003ry,Csanad:2009wc}, such a number density can
be introduced as
\begin{align}
n(x)=n_0\left(\frac{\tau_0}{\tau}\right)^3 \nu(s).
\end{align}
For the momentum distribution of direct photons, this will not be needed, as the only the temperature of the medium
(the strongly interacting plasma) governs the creation of photons, not the density (which however plays an important
role also in the case of hadron creation).

\section{Hadronic and photonic source functions}
The picture used in hydro models is that the pre freeze-out (FO) medium is described by hydrodynamics, and the post freeze-out medium
is that of observed hadrons. The hadronic observables can be extracted from the solution via the phase-space distribution at the
FO. This will correspond to the hadronic final state or source distribution $S(x,p)$. See details about this topic in Ref.~\cite{Csanad:2009wc}.
It is important to see that the same final state can be achieved with different equations of state or initial conditions~\cite{Csanad:2009sk}.
However, as discussed below, the source function of photons is sensitive to the whole time evolution, thus both to initial conditions
and equation of state as well.

In our model the hadronic source distribution of bosons takes the following form~\cite{Csanad:2009wc}:
\begin{align}
S(x,p)d^4x=\mathcal{N}\frac{p_{\mu}\,d^3\Sigma^{\mu}(x)H(\tau)d\tau}{n(x)\exp\left(p_{\mu}u^{\mu}(x)/T(x)\right)-1},
\end{align}
where $\mathcal{N}=g/(2\pi)^3$ (with $g$ being the degeneracy factor), $H(\tau)$ is the proper-time probability distribution of the FO. It is assumed to be a $\delta$ function or a narrow Gaussian centered at the freeze-out proper-time $\tau_0$. Furthermore,
$\mu(x)/T(x)=\ln n(x)$ is the fugacity factor and $d^3 \Sigma_\mu(x)p^\mu$ is the Cooper-Frye
factor~\cite{Cooper:1974mv} describing the flux of the particles, and $d^3 \Sigma_\mu(x)$ is the vector-measure of
the FO hyper-surface. Here the source distribution is normalized such as $\int S(x,p) d^4 x d^3{\bf p}/E = N$,
i.e. one gets the total number of particles $N$ (using $c$=1, $\hbar$=1 units).

For the source function of photon creation we have
\begin{align}\label{e:source}
S(x,p)d^4x = \mathcal{N}\frac{p_{\mu}\,d^3\Sigma^{\mu}(x)dt}{\exp\left(p_{\mu}u^{\mu}(x)/T(x)\right)-1}
= \mathcal{N}\frac{p_{\mu}u^{\mu}}{\exp\left(p_{\mu}u^{\mu}(x)/T(x)\right)-1}\,d^4x
 \end{align}
where $p_{\mu}d^3\Sigma^{\mu}$ is again the Cooper-Frye factor of the emmission hypersurfaces. Similarly to Ref.~\cite{Csanad:2009wc} we assume that the hyper-surfaces
are parallel to $u^\mu$, thus $d^3\Sigma^{\mu}(x) = u^{\mu}d^3x$. This yields then $p_{\mu}u^{\mu}$ which is the energy of the photon
in the co-moving system. Here the simple assumption of a thermal distribution is used, i.e. that the radiation of thermal photons
follows the usual $\propto E/\left[\exp(E/T)-1\right)$ type of behavior. Photon production time has to be also taken into account,
and we assume that photons are created from an initial time $t_i$ until a point sufficiently near the quark-hadron transition.
Note however, that in this case
the photon production rate (integrated over momentum) is not proportional to $T^6$, as expected from some microscopic
models, where
\begin{align}
\textnormal{rate}(A+B\rightarrow X) = n_An_B\left<\sigma_{A+B\rightarrow X} v\right>,
\end{align}
with $n_A$ and $n_B$ being the densities of the two input particles, while $\sigma_{A+B\rightarrow X}$ is the cross-section
of the given process and $v$ is the incoming velocity. Here, however, the densities are proportional to $T^3$ normally, so this
would give a production rate proportional to $T^6$, or even larger if the velocity for example rises with temperature. This
wil bel incorporated in our model as a multiplicative factor in the photon source function, but we will use this only to estimate
a systematic uncertainty of the parameters of our model when comparing to real data. Note that a similar argument is used in UrQMD, where all possible microscopic processes are taken into account, but the measured pion, kaon and proton spectra do not have such a good agreement with this model as the agreement with hydrodynamic models.\label{s:sources}

Experimental observables can then be calculated from the source function, most importantly the invariant
momentum distribution $N_1(p)$ as
\begin{align}\label{e:sn1}
N_1(p)=\int{S(x,p)d^4x},
\end{align}
where again the integration in time is on a definite interval, from the initial time until $\tau_0$. 
To perform this integration we use a second order saddlepoint approximation. In this approximation the point of maximal emissivity is
\begin{align}\label{e:r0}
r_{0,x} = \rho_x t \frac{p_x}{E}\\
r_{0,y} = \rho_y t \frac{p_y}{E}\\
r_{0,z} = \rho_z t \frac{p_z}{E}
\end{align}
while the widths of the particle emitting source are
\begin{align}\label{e:R}
R_x^2 = \rho_x \left( \frac{t}{\tau_0}\right)^{-3/\kappa+2} \tau_0^2 \frac{T_0}{E}\\
R_y^2 = \rho_y \left( \frac{t}{\tau_0}\right)^{-3/\kappa+2} \tau_0^2 \frac{T_0}{E}\\
R_z^2 = \rho_z \left( \frac{t}{\tau_0}\right)^{-3/\kappa+2} \tau_0^2 \frac{T_0}{E}
\end{align}
where we introduced the auxiliary quantities
\begin{align}\label{e:rho} 
\rho_x = \frac{\kappa}{\kappa -3-\kappa\frac{b}{\dot{X_0^2}}}\\
\rho_y = \frac{\kappa}{\kappa -3-\kappa\frac{b}{\dot{Y_0^2}}}\\
\rho_z = \frac{\kappa}{\kappa -3-\kappa\frac{b}{\dot{Z_0^2}}}
\end{align}
where again $\kappa=c_s^{-2}$ is describing the EoS. Note that the source width depend clearly on time, as the system is expanding and so does the photon source.

\section{Thermal photon observables}
The invariant one-particle momentum distribution is defined in eq.~(\ref{e:sn1}), it depends on the three-momentum $p=(p_x,p_y,p_z)$.
We will introduce the $(p_t, \varphi, p_z)$ cylindrical coordinates ($z$ being the beam direction) and use the longitudinal rapidity
$y$ (for which $E\, dy = dp_z$ is true). The PHENIX detector is aligned such that $|y|<0.35$ for photons. Thus, as usual, we restrict our
calculations to $y=0$ from now on (note that in this case $E=p_t$ is true for photons). Our key quantity is then
\begin{align}
\left.N_1(p)\right|_{y=0}=N_1(p_t, \varphi) = \frac{d^2N}{p_t \, dp_t \, d\varphi}
\end{align}
We transform this into two one-dimensional observables, the invariant transverse momentum distribution and the elliptic flow,
similarly to Ref.~\cite{Csanad:2009wc}:
\begin{align}
N_1(p_t, \varphi) = N_1(p_t) \left(1 + 2\sum_{n=1}^{\infty}v_n\cos(n\varphi)\right) 
\end{align}
where $v_2$, the second Fourier component, is the elliptic flow. $N_1(p_t)$ and $v_2(p_t)$ can be calculated from $N_1(p_t,\varphi)$ as
\begin{align}
N_1(p_t) &= \int_0^{2\pi} \! N_1(p) \, d\varphi\\
v_2(p_t) &= \frac{\frac{1}{2\pi}\int_0^{2\pi} \! N_1(p) \cos (2\varphi) \, d\varphi}{N_1(p_t)}
\end{align}

In order to calculate these quantities, we integrate on the space-time coordinates. We also have to integrate on the azimuthal angle $\varphi$.
During that, similarly to Ref.~\cite{Csanad:2009wc} we have to use the modified Bessel functions $I_n$
The result on the invariant transverse momentum distribution depends on the initial and final times. We introduce the variable $\xi=\frac{t}{\tau_0}$
(with $\tau_0$ as the freeze-out time), then in terms of $\xi$ the time integration goes from $i$ to $1$, if $i=\frac{t_i}{\tau_0}$. The result is: 
\begin{align}\label{e:N1ptfinal}
&N_1(p_t) = \sum_{n=0}^{\infty}(2\pi)^{\frac{3}{2}}\sqrt{\rho_x\rho_y\rho_z}\tau_0^4T_0
\left(\frac{p_t}{T_0}\right)^{\frac{3-4\kappa}{3}}\frac{\kappa}{3}\frac{B^n}{A^{n+\frac{4\kappa}{3}-\frac{3}{2}}}\nonumber\\
&\left[  \left(Ca_{0n}+Da_{1n} \right)
\left.\Gamma\left(n + \frac{4\kappa}{3}-\frac{3}{2}, A\frac{p_t}{T_0}\xi^{\frac{3}{\kappa}} \right) \right|_1^i + 
A \frac{\rho_x + \rho_y + \rho_z}{2} \,a_{0n} \left.\Gamma\left(n + \frac{4\kappa}{3}-\frac{5}{2}, A\frac{p_t}{T_0}\xi^{\frac{3}{\kappa}} \right) \right|_1^i \right]\textnormal{.}
\end{align}
where we introduce
\begin{align}\label{e:AB}
A &= 1 - \frac{\rho_x + \rho_y}{4}\\
B &= \frac{\rho_x - \rho_y}{4}\\
C &= 1-\frac{\rho_x+\rho_y}{2}+\frac{\rho_x^2+\rho_y^2}{4}\\
D &= -\frac{\rho_x-\rho_y}{2}+\frac{\rho_x^2-\rho_y^2}{4}
\end{align}
and $a_0$, $a_1$ are the Taylor-coefficients of the first two modified Bessel functions:
\begin{align}
a_{0} = \left(1, 0, \frac{1}{4}, 0, \frac{1}{64}, 0, \frac{1}{2304}, 0, \, ...\right)\\
a_{1} = \left(0, \frac{1}{2}, 0, \frac{1}{16}, 0, \frac{1}{384}, 0, \frac{1}{18432},\, ...\right)
\end{align}
The modified Bessel functions can be expressed then as
\begin{align}
I_0(x) = \sum_{n=0}^{\infty} \! a_{0n}x^n\\
I_1(x) = \sum_{n=0}^{\infty} \! a_{1n}x^n
\end{align}
As the coefficients are strongly decreasing, in real calculations we can restrict ourselves to use only the first two of them, i.e. we can make
the approximation of $I_0(x) = x$ and $I_1(x) = x^2/2$.

For the elliptic flow we get:
\begin{align}\label{e:v2}
&v_2(p_t) = \frac{1}{N_1(p_t)}\sum_{n=0}^{\infty}\frac{(2\pi)^{\frac{3}{2}}\sqrt{\rho_x\rho_y\rho_z}}{N_1(p_t)}\tau_0^4T_0
\frac{\kappa}{3}\left(\frac{p_t}{T_0}\right)^{\frac{3-4\kappa}{3}}\frac{B^n}{A^{n+\frac{4\kappa}{3}-\frac{3}{2}}}\\
&\Bigg\lbrace  \left[ \frac{C-3}{4}(a_{0n}\!+\!a_{2n})+\frac{C-1}{2}a_{1n}\!\right]
\Gamma\left(n\!+\!\frac{4\kappa}{3}\!-\!\frac{3}{2}, A\frac{p_t}{T_0}\xi^{\frac{3}{\kappa}} \right) + 
\frac{\rho_x\!+\!\rho_y\!+\!\rho_z}{2} a_{1n} A
\Gamma\left(n\!+\!\frac{4\kappa}{3}\!-\!\frac{5}{2}, A\frac{p_t}{T_0}\xi^{\frac{3}{\kappa}} \right)
\Bigg\rbrace_1^i\nonumber 
\end{align}
Here $I_2$ is the modified Bessel function, which has the Taylor-coefficients of
$a_2 = \left(0,0,\frac{1}{8}, 0, \frac{1}{96}, 0, \frac{1}{3072}, 0,\, ...\right)$.

\section{Comparison to the measured direct photon spectrum}
Hadronic data were already described with this model in Ref.~\cite{Csanad:2009wc}. Freeze-out properties were thus determined from hadronic fits.
These properties include the expansion rates, the freeze-out proper-time and freeze-out temperature (in the center of the fireball), see Table~\ref{t:param}
for the values of these parameters. We use the parameters of the hadronic fit and leave only the remaining as free
parameters. The free parameters will be $\kappa$ (the equation of state parameter) and $t_i$, the initial time of the evolution.

Similarly to Ref.~\cite{Csanad:2009wc}, we use transverse expansion ($u_t$) and eccentricity ($\epsilon$) instead of $x$ and $y$ direction
expansion rates:
\begin{align}
\frac{1}{u_t^2} = \frac{1}{2}\left( \frac{1}{\dot{X^2}} + \frac{1}{\dot{Y^2}} \right),
\epsilon = \frac{\dot{X^2}-\dot{Y^2}}{\dot{X^2}+\dot{Y^2}}.
\end{align}

\begin{figure}
 \begin{center}
 \includegraphics[width=0.52\textwidth]{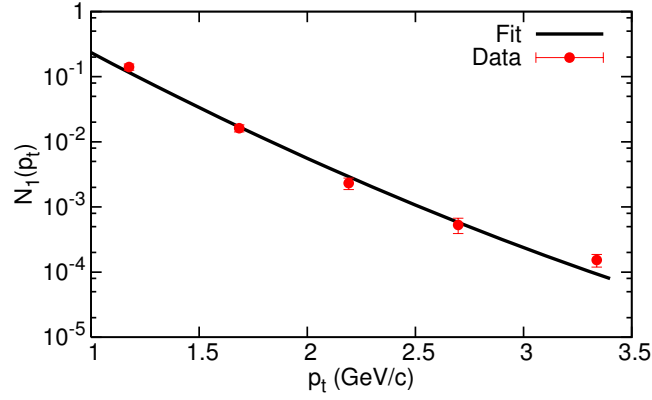}
 \end{center}
 \caption{Invariant transverse momentum of direct photons from our hydro model, normalized to the data. The model
               validity goes until roughly 3 GeV in transverse momentum, the last point on the plot is included in the fit to
               have enough constraints.}\label{f:N1pt}
\end{figure}

We use direct photon data from the PHENIX Collaboration~\cite{Adare:2008fqa}, measured in $\sqrt{s_{NN}}=$200 GeV Au+Au collisions.
For the fits we use the standard Minuit package~\cite{James:1975dr}. Note that we also utilize a normalizing factor to describe the
data. Fit parameters are summarized in Table~\ref{t:param}. The fit itself is shown on Fig.~\ref{f:N1pt}.

\begin{table}
\begin{center}
    \begin{tabular}{ llll }
    \hline\hline
    Parameter &  & Value & Type\ \\hline
    Central FO temperature & $T_0$ & $204$ MeV & fixed \\
    FO proper-time & $\tau_0$ & $7.7$ fm/c & fixed \\
    Eccentricity & $\epsilon$ & $0.34$ & fixed \\
    Transverse expansion & $u_t^2/b$ & $-0.34$ & fixed \\
    Longitudinal expansion & $\dot{Z_0^2}/b$ & $-1.6$ & fixed \\
    Equation of State & $\kappa$ & $7.9 \pm 0.7$ & free \\
    Initial time & $t_i$ & $0 - 0.7$ fm$/c$ & interval \\ \hline 
    Fit property &  & Value & \\ \hline
	  Number of data points & & $5$ &\\    
    Fitted parameters & & $2$&\\ 
    Degrees of freedom  & NDF & $5-2=3$&\\ 
    Chisquare & $\chi^2$ & $7.0$&\\ 
    Confidence level & & $7.2 \%$&\\
    \hline\hline
    \end{tabular}   
\end{center}
\caption{Parameters of the solution, describing the expanding sQGP. The first five are taken from the hadronic fits of Ref.~\cite{Csanad:2009wc} (see more details about the parameters in the reference).
         The EoS parameter $\kappa$ is fitted, while for the initial time we determine an interval of acceptability (with 95\% confidence).}\label{t:param}
\end{table}

\begin{figure}
 \begin{center}
 \includegraphics[width=0.52\textwidth]{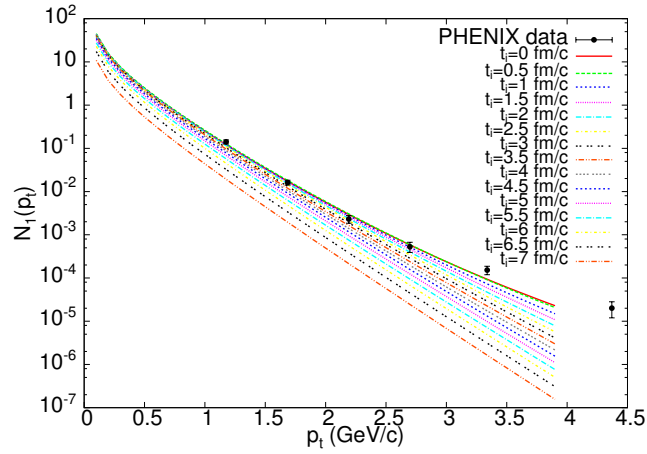}
 \end{center}
 \caption{Invariant transverse momentum of direct photons, calculated with different initial times. This figure
          illustrates the insensitivity on the initial time.}\label{f:N1pt-ti}
\end{figure}

The equation of state result is $\kappa=7.9\pm0.7$, or alternatively, using $\kappa=1/c_s^2$:
\begin{align}
c_s = 0.36\pm0.02
\end{align}
which is in nice agreement with both lattice QCD calculations~\cite{Borsanyi:2010cj} and experimental
results from hadronic data~\cite{Adare:2006ti,Lacey:2006pn}. This represents an average EoS as it may
vary with temperature. There may be solutions with a $\kappa(T)$ function, but for the sake of simplicity we
assumed here an average, fixed $\kappa$. Note however that the spectrum is not very sensitive to the initial
time as in early times the thermal photon
emission is not in the region of the experimental data. See Fig.~\ref{f:N1pt-ti} as an illustration: we plotted the invariant transverse momentum distribution
$N_1(p_t)$ for different values of initial time. Clearly the very first times do not have a large contribution to $N_1(p_t)$ in the desired region. 
Our model does not contain acceleration, it is a Hubble-flow type of model, but the initial acceleration does not play a large role in the 
thermal photon spectrum because of this insensitivity on the initial time. As a result, the fits will neither be sensitive to the exact value
of the initial time -- or otherwise, the initial time $t_i$ will have a very large error. As a workaround, we
determine an ``interval of acceptability'' for $t_i$. The maximum value for $t_i$ within 95\% probability is 0.7 fm/$c$. This can then be used
to determine a lower bound for the initial temperature, using the eq.~\ref{e:temp}. Thus the initial temperature of the fireball (in its center) is:
\begin{align}
T_i > 507\pm12\textnormal{MeV}
\end{align}
at 0.7 fm/$c$. The uncertainty comes from the uncertainty of $\kappa$. This is in accordance with other hydro models as those values are in the 
$300-600$ MeV interval~\cite{Adare:2008fqa}.

As a systematic test, we investigate the changes in the result when using a factor of $(T/T_0)^N$, with $N=0,1,2,3$.
Such a factor would arise if the photon creation can be described by a microscopic process mentioned in Section~\ref{s:sources}.
Such a prefactor causes only minor change in the resulting spectrum, as it is dominated by the exponential factors in it.
However, the equation of state parameter $\kappa$ changes from 7.9 to 6.5 as we increase the exponent in the prefactor. 
It is important also, that the best fit (smallest $\chi^2$) is achieved without the prefactor. 
This test can still be used as a systematic error of our estimate, thus the final result for $c_s$ is then 
\begin{align}
c_s = 0.36\pm0.02_{stat}\pm0.04_{syst}
\end{align}
while for the temperature before 0.7 fm/$c$ we get
\begin{align}
T_i > 507\pm12_{stat}\pm90_{syst}\textnormal{MeV}
\end{align}

\section{Direct photon elliptic flow and correlation radii}
\begin{figure}
 \begin{center}
 \includegraphics[width=0.48\textwidth]{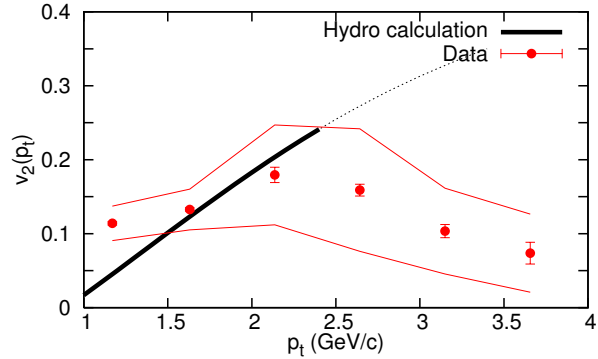}
 \end{center}
 \caption{Thermal photon elliptic flow with parameters from Table~\ref{t:param}, compared
              to PHENIX data of Ref.~\cite{Adare:2011zr}. The eccentricity
               parameter $\epsilon$ is taken from hadronic fits of Ref.~\cite{Csanad:2009wc} as detailed in the
              text.}\label{f:v2pt}
\end{figure}

A measurement of direct photon elliptic flow was also performed recently at PHENIX~\cite{Adare:2011zr}. Using the previously determined fit parameters we can calculate the elliptic flow of direct photons in Au+Au collisions at RHIC.
We compare the elliptic flow $v_2$ from eq.~\ref{e:v2} to data of Ref.~\cite{Adare:2011zr}, using the parameters of Table~\ref{t:param}, with the exception that the
eccentricity parameter $\epsilon$ has to be modified, as the hadronic elliptic flow data of Ref.~\cite{Csanad:2009wc} utilized also
a different $\epsilon$. Due to the low number of points in the desired range, a fit can not be performed here, but we use the
average value $\epsilon$ in case of the two fits of Ref.~\cite{Csanad:2009wc}.
The resulting curve, where  the value $\epsilon=0.59$ is used, is shown on Fig.~\ref{f:v2pt}. Note that already for 1 GeV$/c$
the elliptic flow is very small. It is quite unusual, but $v_2$ is negative for even smaller transverse momenta. This is due to the fact
that in the photon source, the $p_\mu u^\mu$ prefactor (the comoving energy) also oscillates in the azimuthal angle $\varphi$,
as shown in eq.~\ref{e:v2}. After spatial integration the $\varphi$ dependence will look like:
\begin{align}
N_1(\phi)\propto(1-\alpha\cos(2\phi))e^{\beta\cos(2\phi)}
\end{align}
where
\begin{align}
\alpha&=\frac{(\rho_x-1)^2+(\rho_y-1)^2-2}{(\rho_x-1)^2+(\rho_y-1)^2+2+2(\rho_x^2+\rho_y^2+\rho_z^2)/\zeta}\\
\beta &=(\rho_x-\rho_y)\frac{\zeta}{4}
\end{align}
with $\zeta = \frac{T_0}{p_t}\left(\frac{t}{\tau_0}\right)^{3/\kappa}$. This function changes sign in the oscillation at a given point,
depending on $\zeta$. In fact the number of oscillations within the $[0,2\pi]$ interval will be four (in contrast to two oscillations of a regular $\cos(2\varphi)$
function) if $-1<1/\beta-1/\alpha<1$. The two boundaries of this interval can be calculated by solving $\beta=\alpha/(1+\alpha)$ and $\beta=\alpha/(1-\alpha)$
for $p_t$ (which appears in $\zeta$). Exact shape of the elliptic flow depends strongly on input parameters, and also it depends stronger
on initial time than the invariant momentum distribution. Thus an accelerating model might result in a different elliptic flow.

We also calculated Bose-Einstein correlation radii from the above model. As usual, the two-particle correlation function
for identical particles can be calculated from the single particle source function $S$ as
\begin{align}
C_2(q)=1+\lambda\left|\frac{\widetilde S(q)}{\widetilde S(0)}\right|^2.
\end{align}
where $q$ is the momentum difference of the two particles and $\widetilde S(q)$ is the Fourier-transformed of the source $S(x)$.
This transformation can however not be done analytically, so we just calculate the correlation function for the parameter
set used to describe the direct photon spectrum. We analyze the correlation functions to determine the HBT type of
correlation radii. We find the numerically calculated correlation function to have a peak of shape
$C_2(q)=\exp\left|R^2q^2\right|^{\alpha/2}$, where $R^2$ is the matrix of correlation radii, and for $\alpha=2$,
this represent a Gaussian shape. From this shape, we determine the
HBT radii $R_\textnormal{out}$ and $R_\textnormal{side}$ for different average $p_t$ values.

\begin{figure}
 \begin{center}
 \includegraphics[width=0.48\textwidth]{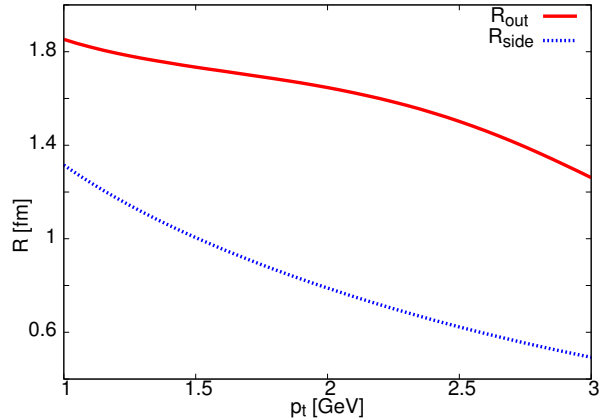}
 \end{center}
 \caption{Thermal photon HBT radii (correlation widths) with example parameters from Table~\ref{t:param}.
         Due to the length of photon production, $R_\textnormal{out}$ is significantly larger than
        $R_\textnormal{side}$. Shape of the correlation functions is nearly Gaussian, optimal values of $\alpha$
        are in the range of 1.7 to 1.9.}\label{f:hbt}
\end{figure}

In case of hadronic HBT, correlation radii in the side and out directions are almost equal, as for the hadronic transition is of
cross-over type (i.e. the transition time is short), see details for example in Ref.~\cite{Csanad:2009wc}. However, in case of
photons, the creation spans the whole evolution of the fireball, thus $R_\textnormal{out}$ will be significantly larger than
$R_\textnormal{side}$.
Indeed this is observed in our model, as shown on Fig.~\ref{f:hbt}. 

\section{Summary}
The medium created in relativistic Au+Au collisions at RHIC turned out to be a strongly interacting, perfect fluid. Observed hadrons
are created at the freeze-out of this fluid, while thermal photons are constantly emitted. Hadrons thus reveal information
about the final state, whereas thermally radiated photons carry information about the whole time evolution. We used
direct photon data of PHENIX to determine the equation of state and the initial temperature of quark gluon plasma. We find that thermal radiation is consistent with the data, and our result on equation of state (average speed of sound) is
$c_s = 0.36\pm0.02_{stat}\pm0.04_{syst}$, and we set a lower bound on the initial temperature of the sQGP to $507\pm12_{stat}\pm90_{syst}$ MeV at $0.7$ fm/$c$. To our knowledge, this is the
first time when these values are extracted from photon and hadron data simultaneously. We also investigated the thermal photon
elliptic flow from this model, which is not incompatible with measurements. We calculated photon HBT radii from the model, and
discovered a significantly larger $R_\textnormal{out}$ than $R_\textnormal{side}$.

\section*{Acknowledgments}
M. Csan\'ad gratefully acknowledges the support of the Bolyai Scholarship of the Hungarian Academy of Sciences and the NK-101438 OTKA grant. The authors also would like to thank T. Cs\"org\H{o} for valuable discussions. M. Csan\'ad thanks the invitation to the WPCF 2011 conference, as interesting discussions took place there as well. In particular, we would like to thank S. Pratt for his important insights in this topic.

\bibliography{../../../master}

\end{document}